\documentclass[letterpaper,11pt]{article}
\usepackage{graphics,graphicx}
\usepackage{longtable}
\usepackage{wrapfig}
\usepackage{lineno}
\usepackage{amsmath,verbatim,amssymb,epsfig,color,lscape}
\usepackage{natbib}
\usepackage{multirow,multicol}
\usepackage[pagestyles]{titlesec}
\usepackage[normalem]{ulem}
\usepackage{bm}
\usepackage{url}
\usepackage{rotating}

\newcommand{\bi}{\begin{itemize}}
\newcommand{\ei}{\end{itemize}}

\makeatletter

\usepackage[outerbars,color]{changebar}
\ifx\pdfoutput\undefined
\else\ifnum\pdfoutput>0
  \usepackage{pdfcolmk}
\fi\fi
\cbcolor{black}

\usepackage[margin=1.0in]{geometry}
\newfont{\rmm}{cmr10 at 11pt}
\rmm

\begin{document}

\title{Social Network Mediation Analysis: a Latent Space Approach}

\author{Haiyan Liu\thanks{
Haiyan Liu, Psychological Sciences, University of California, Merced;
Ick Hoon Jin, Department of Applied Statistics, Yonsei University; 
Zhiyong Zhang, Department of Psychology, University of Notre Dame; 
Ying Yuan, Department of Biostatistics, The University of Texas MD Anderson Cancer Center.
Correspondence concerning this article should be addressed to Haiyan Liu, 
Psychological Sciences, University of California, Merced, 5200 N. Lake Road, Merced, CA 95343. Email: hliu62@ucmerced.edu
}, \ Ick Hoon Jin, \ Zhiyong Zhang, \ and \ Ying Yuan }


\maketitle

\begin{abstract}
A social network comprises both actors and the social connections
among them. Such connections reflect the dependence among social actors,
which is important for individuals' mental health and social development.
In this article, we propose a mediation model with a social network
as a mediator to investigate the potential mediation role of a social
network. In the model, the dependence among actors is accounted by
a few mutually orthogonal latent dimensions which form a social space.
The individuals' positions in such a latent social space directly
involve in the intervention process between an independent variable
and a dependent variable. After showing that all the latent dimensions are equivalent
in terms of their relationship to the social network and the meaning
of each dimension is arbitrary, we propose to measure the whole mediation effect of a network. Although
the positions of individuals in the latent space are not unique, we
rigorously articulate that the proposed network mediation effect is
still well-defined. We use a Bayesian method to estimate the model
and evaluate its performance through an extensive simulation study
under representative conditions. The usefulness of the network mediation
model is demonstrated through an application to a college
friendship network.
\end{abstract}

\section{Introduction}

Network analysis is an interdisciplinary research topic of mathematics,
statistics, and computer sciences \citep{wasserman1994social,schmittmann2013deconstructing,epskamp2017generalized}.
It has been adopted in diverse fields to address different research
interests \citep{Grunspan2014}. Researchers have been working on
social networks from different perspectives \citep{carrington2005models}.
Graph theory is often used by mathematicians to examine the network
structure \citep{Newman2002}. Different modeling frameworks and algorithms
are developed by computer scientists and statisticians to detect and
understand network communities \citep{zhao2012consistency,Yang2013}.
Probability and statistical models with social networks as dependent
variables are built to understand the dependence of actors in a network
\citep{snijders2011statistical}. Representative models include but
are not limited to block models \citep{airoldi2008mixed,anderson1992building,choi2012stochastic,holland1983stochastic,Nordlund2019,Sweet2018},
exponential random graph models \citep{Anderson1999,Lusher2013,Snijders2002},
and latent space models \citep{Hoff2002,paul2016,liu2018}.

Network analysis also has a long history in psychology and sociology,
which is called \emph{social network analysis} (SNA) that mainly focuses
on social relations among actors \citep{wasserman1994social,Westaby2014}.
Social relations have been traditionally studied in psychology and
sociology and it has been found that (1) social relations influence
people's subjective well-being over the life course \citep{House1988,Gurung1997,McCamish-Svensson1999,Seeman2001,umberson2010social,Cacioppo2014};
(2) close social relations such as marriage and friendship predict
late-life health and well-being \citep{Waldinger2015}; (3) social
relations with peers also influence individuals' health behaviors
\citep{Broman1993,umberson2010social}; and (4) social cohesion can
ease smoking cessation \citep{reitzel2012relation}. Because of the
tremendous importance of social connections, explaining their formation
is of enormous interests to researchers. Researchers are also highly
interested in actors attributes in social networks. Actor attributes
such as personalities are found to be closely related to close relations
like friendship and marriage \citep{Asendorpf1998,mccrae2008personality,Harris2016a,liu2018}.
The similarity in academic achievement predicts the friendship ties
among students \citep{Flashman2012}.

A social network contains data on both social relations and actor
attributes \citep{wasserman1994social}. It provides a platform for
researchers to study traditional research questions, such as the association
between social relations and actor characteristics, from a network
perspective \citep{liu2018}. To offer a motivation example, we are
interested in how students' attributes (e.g., gender) influence their
friendship network and, in turn, affects their health behaviors (e.g.,
smoking) in our empirical study. Such a research question can be addressed
by mediation analysis.

Mediation analysis is a common framework for statistical analysis
of the causal mechanism of the observed relationship between an independent
variable and a dependent variable \citep{baron1986moderator,hayes2009beyond}.
It is a popular research discipline in epidemiology, psychology, sociology,
and related fields \citep[e.g.,][]{fritz2007required,richiardi2013mediation}.
The object of mediation analysis is to determine whether the association
between two variables is due, wholly or in part, to a third variable
($M$) that transmits the effect of an independent variable (i.e.,
$X$) on the dependent variable (i.e., $Y$) \citep{mackinnon2007mediation,mackinnon2012introduction}.
\textcolor{blue}{In social sciences, regression-based approaches are
commonly used to investigate the relationship between the mediator
(i.e., a third variable), the independent variable, and the dependent variable.
However, the significant indirect effect obtained using the tri-variate
regression system is not necessarily the ``mediation'' effect. The significant indirect effect could also the potential confounding effect \citep{mackinnon2000equivalence,Sweet2019,valeri2013mediation}.
To draw causal inference using the regression approach, one must know
that it is a causal path from $X$ to $M$ to $Y$ based on theory
\citep{imai2010general,Pearl2014}.}

Over the years, mediation analysis has achieved significant progress
along two equally important methodological lines: more reliable approaches
for hypothesis testing of the mediation effects and new models for
conducting mediation analysis under different contexts \citep[e.g.,][]{Yu2018,zhang2017}.
For instance, multilevel mediation models were proposed to measure
the mediation effects with clustered data \citep[e.g.,][]{Kenny2003}
and longitudinal models were developed to do mediation analysis with
longitudinal or time series data \citep[e.g.,][]{Cheong2003,Cole2003}. And very recently, Zhang and Philips \citeyearpar{Zhang2018}
proposed a three-level model to study the longitudinal mediation effect
in nested data. \textcolor{blue}{
\citet{Yu2018} proposed a nonlinear model for multiple mediation
analysis.} Many approaches have been proposed to test
the mediation effects in the frequentist framework \citep{mackinnon2002comparison}.
The \emph{joint significance test} approach \citep{mackinnon2004confidence}
and the \emph{bootstrap} approach \citep{preacher2008asymptotic}
are recommended over the \emph{Sobel test} because they have higher
power and more accurate Type I error rates \citep{mackinnon2002comparison,preacher2012advantages}.
Bayesian methods are also used in mediation analysis and they are
expanding their applications in this field \citep{yuan2009bayesian,wang2011estimating,enders2013bayesian,wang2015moderated,miovcevic2018tutorial}.
Bayesian methods can lead to more efficient parameter estimates with
the prior information. In addition, Bayesian inference of the mediation
effect is based on its posterior distribution, but not the distribution
of the test statistics. Thus, it is especially useful with small sample
sizes when the asymptotic distributions of test statistics are not
available \citep{yuan2009bayesian}.

\textcolor{blue}{To address the substantive interests in the social
network mediation effect, researchers have started to build models
and to develop estimation methods for the mediation analysis with social
network data. Very recently, \citet{Sweet2019} proposed a model for estimating the mediation effect of networks using the stochastic block models. \citet{Sweet2019} aimed to investigate how
the network structure mediates the effect of an independent variable
on the dependent variable. It treated each network as an observation
and the actual mediator was a ``statistic'' summarizing the information
of the entire network. The study units were networks and researchers
should have observations on a sample of networks to use that model. In a study we conducted, we investigated how to measure the mediation effect with a single network (reference masked for the review purpose). However, the purpose of that study was to provide a tutorial on how to apply network mediation analysis. There were no theoretical justification and simulation evaluation of the method.}

\textcolor{blue}{In the current study, we focus on actors' behaviors and are interested in how ``social positions of actors'' in a bounded social network mediate
the effect of an independent variable on a dependent variable. Therefore, our model will focus on social actors in a network and it is
an actor-level analysis of social networks \citep{Clifton2017}. However, the distinct characteristics of network data pose unique
challenges to actor-level mediation analysis.}

Social network data are often represented by an adjacency matrix.
For a network with $N$ actors, its adjacency matrix ${\bf M}$ is
a square matrix of dimension $N$ by $N$. For a pair of actors $(i,j)$
in the network, the element $m_{ij}$ at the node of row $i$ and
column $j$ represents the social relation from actor $i$ to actor
$j$. It takes values either 0 or 1 indicating the absence or presence
of a certain social relation in a binary network. Each row of ${\bf M}$
stores the social relation of the row actor with the others within
the network. The diagonal elements of ${\bf M}$ are 0 unless each
individual can have a self-connection. When the social relation is
undirected, the relation from actor $i$ to actor $j$ is the same
as from actor $j$ to actor $i$, so that the adjacency matrix of
a network is symmetric.

The unique format of social network data poses challenges to statistical
inference. First, network data are high-dimensional. The smallest
unit in a social network is a dyad, which is a pair of actors and
the possible relations between them. Given an undirected binary social
network (i.e., a network with undirected and binary social relations)
with $N$ actors, each actor appears in $N-1$ dyads. In total, there
are $\binom{N}{2}$, i.e., $\frac{N(N-1)}{2}$, dyads in an undirected
social network, which grows with a speed of $O(N^{2})$ as the number
of actors $N$ increases. Therefore, there is a discrepancy
between the dimensions of network data and other actor attributes
data. Second, dyads in a network are not independent of each other.
For example, two dyads sharing a common actor depend on each other.
Testing and explaining the social network dependence has become a
popular research area \citep[e.g.,][]{liu2018,Su2019}. The high-dimensionality
and the dependence of social network data violate the assumptions
of most commonly used statistical modeling tools where independent
observations are required and the dependent variables should have
the same dimension as the independent variables. Therefore, traditional
mediation techniques such as the regression analysis are not directly
applicable in studying the mediation effect of a social network.

The goal of this study is thus to fill the current gap in the literature
by proposing a mediation model and developing an estimation method
to measure the mediation effect of individuals' social positions in
a social network. To deal with the high-dimensionality and dependence
of social network data, we find a low-dimensional representation of
a social network using a latent space model \citep[e.g.,][]{Handcock2007,Hoff2002,Krivitsky2009,Sewell2015}.
Each actor holds a position in such a lower-dimensional space and
the actors' latent positions are then used in the mediation analysis.
Latent space modeling maps actors in a network into an unknown latent
social space with a few dimensions. The distance of two actors in
the latent space predicts the existence of connections between them.
Therefore, the latent positions of actors largely explain the observed
network structure. In our mediation model, the latent positions are
directly involved in the intervention process between the independent
and dependent variables as actual ``mediators''. Hence, our model
contains both a measurement model explaining an observed network using
actors' positions in the latent social space and a mediation model
with the latent positions as actual mediators.

\textcolor{blue}{Because all latent dimensions are equivalent to each
other in terms of their relations with the observed social network,
we cannot label them without unique information on each dimension.
The effect along a single dimension is of limited practical interest.
However, individuals' positions in the latent space represent the
``social position'' of them in the social network. As a consequence,
the positions as a whole in the entire latent space are meaningful and informative. We thus focus on
the overall indirect effect of all dimensions of the latent space.} Although the latent positions of actors are not unique for a given
network \citep{Hoff2002} \textcolor{blue}{without additional restrictions}
due to the invariance property of Euclidean spaces, we will rigorously
articulate that the newly proposed network mediation effect is still
well-defined given the dimension of the latent space. To obtain an
estimate of the network mediation effect, we will adopt a Bayesian estimation
(BE) method, which is also used in the simple mediation analysis \citep{yuan2009bayesian}.

The rest of this article is structured as follows. First, we briefly
introduce a mediation model in the context of the simple mediation
analysis. Next, we present our network mediation model and its assumptions.
We then define the network mediation effect and show that it is well-defined.
After that, we explain the settings in a Bayesian estimation method
for estimating the network mediation effect. A simulation study is
conducted to evaluate the performance of the Bayesian estimation method
in estimating the network mediation model. A detailed empirical example
is used to demonstrate the application of this model. We conclude
the study with discussions on the current development and future directions.

\section{A Brief Introduction to Mediation Analysis}

In this section, we briefly introduce the simple mediation analysis.
The basic mediation framework involves a three-variable system in
which an independent variable $X$ predicts a dependent variable $Y$
via regression models \citep{baron1986moderator}, which is demonstrated
by the diagrams in Figure \ref{fig:mediation1}. The diagram on the
top panel of Figure \ref{fig:mediation1} portrays the total relation
between the independent and dependent variables and the regression
equation is as follows: 
\begin{equation}
\mbox{Model 1:}\quad Y_{i}=i_{1}+cX_{i}+\varepsilon_{i,1},\label{eq: model 1}
\end{equation}
where the coefficient $c$ is the \emph{total effect} of the independent
variable $X$ on the dependent variable $Y$ without considering a
third variable, $i_{1}$ is the intercept of the model and $\varepsilon_{i,1}$
is the error term associated with case $i$. The diagram on the bottom
panel of Figure \ref{fig:mediation1} is a mediation model with the
variable $M$ as the mediator or intervening variable. To study the
indirect effect of $X$ on $Y$ through a mediator variable $M$,
one needs to regress $M$ on $X$ and then $Y$ on both $X$ and $M$,
\begin{align}
 & \mbox{Model 2:}\quad M_{i}=i_{2}+aX_{i}+\varepsilon_{i,2}\label{eq:model 2}\\
 & \mbox{Model 3:}\quad Y_{i}=i_{3}+bM_{i}+c'X_{i}+\varepsilon_{i,3},\label{eq:model 3}
\end{align}
where $i_{2}$ and $i_{3}$ are the intercepts of the two regression
models. The parameter $a$ is the coefficient of the relation between
$X$ and $M$, $b$ is the coefficient relating the mediator $M$
to $Y$ while controlling $X$, and $c'$ is the coefficient quantifying
the relationship between $X$ and $Y$ while controlling $M$. The
two terms $\varepsilon_{i,2}$ and $\varepsilon_{i,3}$ are errors
associated with case $i$ in these two models. 

\begin{figure}
\begin{centering}
\setlength{\unitlength}{1mm} \begin{picture}({85, 20}) \put(
0,10){\framebox(25,10){\textbf{X}}} \put(60,10){\framebox(25,10){\textbf{Y}}}
\put(25,15){\vector(1,0){35}} \put(42,18){$c$} \put(10,
5) {a) Path diagram for the regression model.} \end{picture} \\
 \begin{picture}(85,60) \put( 0,10){\framebox(25,10){\textbf{X}}}
\put(60,10){\framebox(25,10){\textbf{Y}}} \put(30,45){\framebox(25,10){\textbf{M}}}
\put(25,15){\vector(1,0){35}} \put(11,20){\vector(2,3){19}}
\put(55,49){\vector(2,-3){19}} \put(17,34){a} \put(66,34){b}
\put(42,18){c'} \put(-25, 5){b) The simple mediation model with
$M$ as a mediator of the effect of $X$ on $Y$.} \end{picture} 
\par\end{centering}
\caption{Path diagrams for the regression model and the mediation model. \label{fig:mediation1}}
\end{figure}
{[}Insert Figure 1 here{]}\\The \emph{indirect effect} is the estimate
of the reduction in the predictor effect on the outcome variable when
the mediator is included in the model, that is $\hat{c}-\hat{c}'$
given a sample. In general, it holds that $\hat{c}-\hat{c}'=\hat{a}\times\hat{b}$
when the three variables are linearly related to each other \citep{mackinnon1995simulation}.
The rationale behind this method is that the mediation effect depends
on the extent to which the predictor changes the mediator, represented
by the coefficient $a$ and the extent to which the mediator affects
the outcome variable, represented by the coefficient $b$.

\textcolor{blue}{The standard regression approach is commonly used
in social sciences to study the mediation effect. However, the indirect
effect $\hat{a}\times\hat{b}$ obtained using the above tri-variable
regression system does not necessary indicate a causal mediation effect \citep{mackinnon2000equivalence,VanderWeele2009,VanderWeele2015}
without assumptions. To make the tri-variable regression system to
be a mediation model, the path from $X$ to $M$ to $Y$ should be
causal. Specifically, there should be no unmeasured confounders for
$X$ to $Y$ relationship, the $X$ to $M$ relationship, and
the $M$ to $Y$ relationship \citep{imai2010general,VanderWeele2009,VanderWeele2015}.
In addition, if there are confounders for $M$ and $Y$, they should
not be affected by the independent variable $X$. More detailed discussions
on causal mediation analysis can be found in the work by \citet{VanderWeele2015}.}

\section{Proposed Network Mediation Model}

\textcolor{blue}{In the current study, we will develop an actor-level
network mediation model to investigate how actors' social positions
mediate the effect of an independent variable on the dependent variable.}
Because the dyadic variable of a network is of a size $O(N^{2})$
and the size of actor attribute data is $N$ with $N$ being the number
of actors, it is not feasible to directly use the network data in
the analysis. Instead, we will use the latent space modeling approach
to map the actors into \textcolor{blue}{a low-dimension latent social
spac}e, which represents a given social network using only a few dimensions
\citep{Hoff2002}. The positions of actors in the latent space will
act as actual mediators and the indirect effect through them will
be estimated. The newly proposed model thus consists of two parts:
a latent space model explaining the social connections in an observed
social network using a few latent dimensions and a mediation model
with actors' social positions acting as mediators. For the ease of
presentation, we restrict the current study in the context of an undirected
binary social network. So in the adjacency matrix ${\bf M}$, the
element $m_{ij}$ on the node $(i,j)$ takes a value 1
if actors $i$ and $j$ are connected and 0
if they are not. Because the relation between the two actors is undirected,
$m_{ij}$ and $m_{ji}$ are equal.

\subsection{Lower-dimension representation of social networks}

To deal with the challenge posed by the high-dimensional format of
social network data, we adopt the latent space modeling approach to
find its lower dimension representation \citep{Hoff2002}. A latent
space model assumes that each actor holds a position in a $D$-dimensional
Euclidean space $\mathbb{R}^{D}$ with $D$ being a natural number
and much less than the number of actors $N$. The axes of the Euclidean
space represent actors' latent characteristics influencing the formation
of connections between actors. The relative positions of actors in
the latent space describe their ``closeness'' and predict the social
relations in a network through a logistic function,

\begin{align}
\text{Latent space model: } & \begin{cases}
m_{ij} & \sim\text{Bernoulli}(p_{ij})\\
\text{logit}(p_{ij}) & =\alpha-|{\bf z}_{i}-{\bf z}_{j}|,
\end{cases}\label{eq:network mediation 1}
\end{align}
where ${\bf z}_{i}=(z_{i1},\cdots,z_{iD})^{t}$ and ${\bf z}_{j}=(z_{j1},\cdots,z_{jD})^{t}$
are the latent positions of actors $i$ and $j$, which represent
their locations in the underlying social space; and $|\textbf{z}_{i}-\textbf{z}_{j}|$
is the Euclidean distance between actors $i$ and $j$, such that
\begin{equation}
|\textbf{z}_{i}-\textbf{z}_{j}|=\sqrt{\sum_{d=1}^{D}(z_{id}-z_{jd})^{2}}.\label{eq:Euclidean distance}
\end{equation}
As in most statistical models, local independence is assumed in a
latent space model. Specifically, the dyads are conditionally independent
of each other given latent positions, 
\begin{equation}
P(m_{ik}=1,m_{jk}=1|{\bf z}_{i},{\bf z}_{j},{\bf z}_{k})=P(m_{ik}=1|{\bf z}_{i},{\bf z}_{k})P(m_{jk}=1|{\bf z}_{j},{\bf z}_{k}),
\end{equation}
which indicates that the dependence among dyads is totally explained
by the actors' latent positions.

\textcolor{blue}{As discussed in by \citet{Hoff2002}, the latent
space modeling is a non-linear model-based multidimensional scaling
technique. It aims to find a few dimensions that could explain sufficient
``similarity'' (i.e., connections) among actors within a network.
Dimensions extracted from social networks are latent characteristics
that explain the formation of social relations. Because the use of
the Euclidean distance (Equation \ref{eq:Euclidean distance}),
all dimensions are equivalent to each other in terms of the relationships
with the manifest social network. Therefore, we cannot label them
without unique information on each dimension, which is used to be
an issue also in factor analysis \citep{Cattell1952}.}

A potential question raised for the application of the latent space
model (Equation \ref{eq:network mediation 1}) in substantive research
is how many dimensions the latent space should have. When choosing
the dimensions, we should consider both the model adequacy and complexity.
The accuracy of prediction is improved when the latent space has more
dimensions but with a cost of more parameters. A model fit index with
a penalty term for model complexity could be used. In our empirical
study, we will demonstrate how to determine the dimension of the latent
space using the Bayesian Information Criterion \citep[BIC,][]{Schwarz1978}.

\subsection{Mediation model}

Actors' positions in the latent social space (i.e., latent Euclidean
space) are the actors' underlying attributes explaining the observed
network. A partial/full effect of the independent variable $X$ on
the outcome variable $Y$ \textcolor{blue}{can be mediated} by the
latent positions forming the social network.

The second part of our model is thus a mediation model with multiple
mediators, 
\begin{equation}
\text{Mediation model: }\begin{cases}
{\bf z}_{i} & =\bm{i}_{1}+\bm{a}X_{i}+\bm{\varepsilon}_{i1}\\
Y_{i} & =i_{2}+\bm{b}^{t}{\bf z}_{i}+c'X_{i}+\varepsilon_{i2},
\end{cases}\label{eq:network mediation 2}
\end{equation}
where $(\cdot)^{t}$ is the transpose of a vector or matrix, ${\bf z}_{i}=({z}_{i,1},\cdots,{z}_{i,D})^{t}$
is the latent position of actor $i$ in the Euclidean space $\mathbb{R}^{D}$,
$\bm{i}_{1}=(i_{1,1},i_{1,2},\cdots,i_{1,D})^{t}$ is the column vector
of intercepts of the regression model from the independent variable
to latent positions, the scalar $i_{2}$ is the intercept of the regression
model from latent positions to the outcome variable $Y$. In Equation
\eqref{eq:network mediation 2}, $\bm{a}$ and $\bm{b}$ are both
column vectors of slope parameters from the independent variable $X$
to mediators ${\bf {z}}$ and from ${\bf z}$ to the outcome variable
$Y$. Because the latent positions are in the $D$-dimensional Euclidean
space, both $\bm{a}$ and $\bm{b}$ are of length $D$, i.e, $\bm{a}=(a_{1},a_{2},\cdots,a_{D})^{t}$
and $\bm{b}=(b_{1},b_{2},\cdots,b_{D})^{t}$. In the model, $c'$
is the direct effect of $X$ on $Y$ after controlling latent positions.
$\bm{\varepsilon}_{i1}$ is a $D\times1$ vector of residuals of case
$i$ when mediators are regressed on the independent variable and
$\varepsilon_{i2}$ is the residual of case $i$ when the outcome
variable is regressed on both the independent variable and mediators. 

\begin{figure}[!h]
\begin{centering}
\includegraphics[scale=0.4]{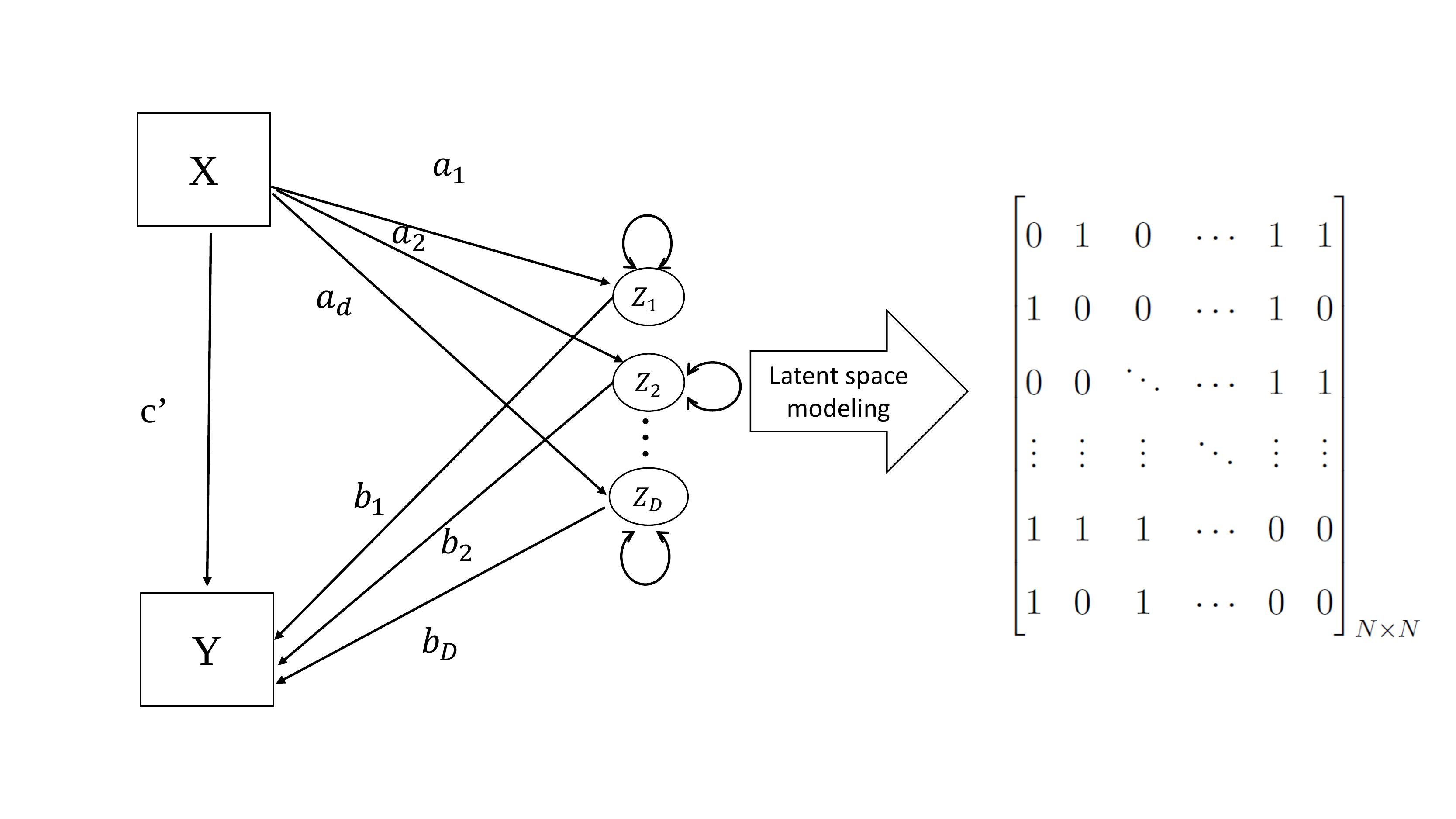} 
\par\end{centering}
\caption{Path diagram of a network mediation model with $D$ dimensions in
the latent space }
\label{diagmediation} 
\end{figure}

\textcolor{blue}{Following the tradition in classical mediation analysis,
the residuals $\varepsilon_{i1}$ and $\varepsilon_{i2}$ are assumed
to be independent across individuals. The residual vector $\bm{\varepsilon}_{1}$
is assumed to follow a multivariate normal distribution with mean
${\bf 0}$ and a diagonal covariance matrix $var(\bm{\varepsilon_{1}})=\mbox{diag}(\sigma_{1,d}^{2},d=1,2,\cdots,D)$.
Therefore, the latent positions ${\bf z}_{i}$ are conditionally independent
with each other given $X$. And the residual $\varepsilon_{2}$ also
follows a normal distribution with mean 0 and variance $var(\varepsilon_{2})=\sigma_{2}^{2}$.
We also assumed that $\bm{\varepsilon_{1}}$ and $\varepsilon_{2}$
are independent with each other.} Figure \ref{diagmediation} is the
path diagram of the model defined by Equations \eqref{eq:network mediation 1}
and \eqref{eq:network mediation 2}. 
The proposed mediation model comprises both Equations \eqref{eq:network mediation 1}
and \eqref{eq:network mediation 2}. The former is used to reduce
the dimension of the network data and the latter is used to bridge
the independent and dependent variables using actors' latent positions.
In the proposed model, a social network manifests the social positions
of its actors. The Euclidean distance of two actors in the latent
space predicts how likely they are connected in the manifest social
world. The manifest social network is the source to provide information
on latent positions. \textcolor{blue}{Because of the use of Euclidean
distance in the link function as in Equation \eqref{eq:network mediation 1},
all the $D$ latent dimensions have a similar relationship with the
social network, and thus, we are not able to label them because of
the lack of unique information for each dimension. A potential solution
to this problem is to introduce indicators for each latent dimension
as done for the factor identification in factor analysis \citep{Cattell1952}.
In the current setting of the network mediation analysis, only linear
models with no directional paths across latent dimensions are considered. }

\textcolor{blue}{The proposed model in Equation \eqref{eq:network mediation 2} can be used to investigate and test the mediation effect of individuals'
social positions within a network. Similar to the classical mediation model \citep{VanderWeele2009,VanderWeele2015},
one needs to show that there are no unmeasured confounders for the
X to ${\bf z}$ relationship, the ${\bf z}$ to $Y$ relationship,
and the $X$ to $Y$ relationship to conclude causal mediation. In the current model specification,
we also assume that there are no directional paths across latent dimensions.
This condition implies that there should be no confounders for the
mediator-outcome relationship that are affected by the independent
variable X. All those assumptions are similar to those for the classical
mediation model \citep{imai2010general,vanderweele2010odds,VanderWeele2015}. }

\textcolor{blue}{Although both \citet{Sweet2019} and our newly proposed
model are for the network mediation analysis, the two modeling frameworks
are fundamentally different. First, the study units of the model proposed
by Sweet (2019) were networks. The data consisted of observations on
a sample of social networks. The model focused on how the exposure
variable changed the entire network characteristics and thus impacted
the outcome variables. It was a network-level mediation model. The
actual mediator was a statistic of the entire
network. While in our model, the study units are actors
within a bounded social network and the data contain both actors'
characteristics and their social relations with others within the
same network. The proposed model can be used to evaluate how individuals'
characteristic influences their social relations with others (described
by the social positions) and how social relations with others, in
turn, influence their behaviors. Therefore, the focus of the newly
proposed model is on actors, but not the entire network. Moreover,
the two modeling frameworks use different approaches to address the
high-dimensionality of social network data. In \citet{Sweet2019},
a parameter $\gamma$ from the stochastic block model replaced the
position of a network in the modeling. In our propose model, we use
a latent space approach \citep{Hoff2002} to find a low-dimension
representation of a network. The actors' positions in the latent space
are actual mediators. }

The newly proposed mediation model is also the study by \citet{liu2018} in both study objectives and model
formulations. First, the analysis units of the study by \citet{liu2018}
are dyads. It aims to predict the formation of social relations using
actors' personality traits and some other attributes. The latent space
is thus replaced by a latent personality space. Second, latent personality
traits in the study by \citet{liu2018} are measured by self-reported
data on personalities. Consequently, the latent personality dimensions
can be labeled based on the items used to measure them. The object
of our newly proposed mediation model is to study how actors' attributes
influence their dependence among each other within a social network
and how such dependence, in turn, affects actors' social behaviors.
A social network in the mediation model acts thus as both a dependent
variable and an independent variable. Furthermore, there is no additional
information available on the latent space besides the observed social
network, which \textcolor{blue}{poses additional challenges on the model
parameter estimation in social network mediation analysis.}

\section{Network Mediation Effect}

Given the dimension $D$ of a latent space, the latent position variables
describing the the social positions of actors are ${\bf z}=(z_{1},z_{2},\cdots,z_{D})^{t}$.
The coefficients from the independent variable to the latent position
and from the latent position to the outcome variable are $\bm{a}=(a_{1},a_{2},\cdots,a_{D})^{t}$
and $\bm{b}=(b_{1},b_{2},\cdots,b_{D})^{t}$, both of which are column
vectors of $D$ parameters. The direct effect from the independent
variable to the dependent variable is denoted by the parameter $c'$.

\textcolor{blue}{Because there are multiple latent variables
that jointly mediate the relationship between the independent and
dependent variables} as in Figure \ref{diagmediation}, a primary
question to ask is what kind of effects are well-defined and testable.
In the newly proposed mediation model, there are potentially three
types of effects of interest. The first one is the product of the
coefficients along each dimension, i.e., $a_{d}b_{d}$, which can
be used to describe \emph{the indirect effect of a single dimension.}
The second one is\emph{ the direct effect} of an independent variable
$X$ on a dependent variable $Y$, represented by $c'$ (Figure \ref{diagmediation}).
The third effect that can be obtained using the proposed model formulation
(Equation \ref{eq:network mediation 1} and \ref{eq:network mediation 2})
is \emph{the indirect effect of a social network} as a whole. 

\textcolor{blue}{In the newly proposed mediation model, the $D$ latent
dimensions are measured by the same social network and they are conditionally
independent given $X$. Based on the model assumptions, the variables
are linearly related and there are no interactions among latent dimensions.
Therefore, we define the network mediation effect as
\begin{equation}
med={\bm{a}}^{t}{\bm{b}}=\sum_{d=1}^{D}a_{d}b_{d},\label{eq:indirect}
\end{equation}
which quantifies the number of units change on $Y$ for a unit change
on $X$ that goes through the latent positions ${\bf z}$.}

As discussed by \citet{Hoff2002}, the latent distance $d_{ij}=|{\bf z}_{i}-{\bf z}_{j}|$
directly predicts dyads in a social network. The Euclidean distance
is invariant to the operations of \emph{translation}, \emph{rotation},
and \emph{reflection} of the latent space. Specifically, if a set
of actor positions $\{{\bf z}_{i}\}_{i=1}^{N}$ are the optimal positions
predicting a social network, then another set of positions $\{{\bf z}_{i}^{*}\}_{i=1}^{N}$
translated, rotated, or reflected from $\{{\bf z}_{i}\}_{i=1}^{N}$
are also optimal because their pairwise distances are the same as
their counterparts computed using $\{{\bf z}_{i}\}_{i=1}^{N}$. However,
regression coefficients in the network mediation model (Equation \ref{eq:network mediation 2})
are not necessarily the same, when a different set of latent positions
are used. Consequently, it is still unclear whether the three types
of effects are uniquely determined or not. In the following, the identification
issues of the three types of effects are discussed one by one.

\subsection{Indirect effect in each dimension}

In the proposed mediation model, the indirect effect of each latent
dimension, i.e, $a_{d}b_{d}$, is not well-defined. When the latent
positions are rotated around origin clockwise $90$ degrees, the coordinate
of an actor coordinate in the latent space shifts to lefts. The positions
of actors becomes ${\bf z}_{i}^{*}=(z_{i,2},z_{i,3},\cdots,z_{i,D},z_{i,1})^{t}$.
Therefore, the regression coefficients using $\{{\bf z}_{i}^{*}\}_{i=1}^{N}$
are ${\bm{a}}^{*}=(a_{2},a_{3},\cdots,a_{D},a_{1})^{t}$ and ${\bm{b}}^{*}=(b_{2},b_{3},\cdots,b_{D},b_{1})$.
Thus the indirect effect along each individual dimension varies when
the latent positions are rotated around the origin.

In addition, all latent dimensions are measured by the same network,
they are thus equivalent to each other in terms of the relations to
the observed network. As a result, it is impossible to name the latent
dimension and not meaningful to study the indirect effect of an individual
dimension.

\subsection{Indirect effect of a social network and direct effect}

Nonetheless, the mediation effect
of a social network as a whole can be well-defined. In our model (Equation
\ref{eq:network mediation 1} and \ref{eq:network mediation 2}),
the variables $X$ and ${\bf z}$ and $Y$ are linearly related. More
over, there is no interaction between latent dimensions. In
the current study, the network mediation effect is based on the inner
product of coefficients, 
\begin{equation}
med={\bm{a}}^{t}{\bm{b}}=\sum_{d=1}^{D}a_{d}b_{d}.\label{eq: nmediatineff}
\end{equation}
When $D=1$, there is only one latent variable underlying the social
network. The mediation effect is $a_{1}b_{1}$, which is similar to
the conventional mediation analysis with a latent mediator. When $D>1$,
there are multiple latent variables. The mediation effect defined
in Equation \eqref{eq: nmediatineff} is the change on $Y$ resulted
from a unit change on $X$ that passes through the social network.

Because there are multiple sets of latent positions maximizing the
likelihood in predicting the social network, it is yet unknown whether
the network mediation effect in Equation \eqref{eq: nmediatineff}
is uniquely defined or not for a given social network. In the following,
we are going to show that the quantity defined in Equation \eqref{eq: nmediatineff}
is invariant to \emph{translation}, \emph{rotation}, and \emph{reflection}
given the latent dimension $D$.

\subsubsection{Translation}

The operation of \emph{translation} is to slide a position to a new
position. For instance, moving a position left or right, up or down
are cases of translation. Given a constant column vector ${\bf t}=(t_{1},t_{2},\cdots,t_{D})^{t}$,
let $T_{{\bf t}}$ be a translation operator such that 
\begin{equation}
{\bf z}_{i}^{*}=T_{{\bf t}}({\bf z}_{i})={\bf z}_{i}+{\bf t},\text{for any actor }i=1,\cdots,N,
\end{equation}
where ${\bf z}_{i}^{*}$ is the ``new'' position of actor $i$ after
translation, and ${\bf t}=(t_{1},t_{2},\cdots,t_{D})^{t}$ is the
difference between the two positions before and after the transition.
When fitting the mediation model (Equation \ref{eq:network mediation 2})
to the new latent positions, we have 
\begin{equation}
\begin{cases}
{\bf z}_{i}^{*} & =\bm{i}_{1}^{*}+\bm{a}^{*}X_{i}+\bm{\varepsilon}_{i1}^{*}\\
Y_{i} & =i_{2}^{*}+(\bm{b}^{*})^{t}{\bf z}_{i}^{*}+c'^{*}X_{i}+\varepsilon_{i2}^{*}.
\end{cases}\label{eq:translation1}
\end{equation}
Because ${\bf z}_{i}^{*}={\bf z}_{i}+{\bf t}$, then the above regression
models become 
\begin{equation}
\begin{cases}
{\bf z}_{i}+{\bf t} & =\bm{i}_{1}^{*}+\bm{a}^{*}X_{i}+\bm{\varepsilon}_{i1}^{*}\\
Y_{i} & =i_{2}^{*}+(\bm{b}^{*})^{t}({\bf z}_{i}+{\bf t})+c'^{*}X_{i}+\varepsilon_{i2}^{*}.
\end{cases}\label{eq:translation2}
\end{equation}
By comparing the two sets of coefficients in Equations \eqref{eq:network mediation 2}
and \eqref{eq:translation2}, it is clear that 
\begin{equation}
{\bm{a}}^{*}={\bm{a}}\text{ and }{\bm{b}}^{*}={\bm{b}}\text{ and }c'^{*}=c.
\end{equation}
Therefore, $med^{*}={\bm{a}^{*}}^{t}\bm{b}^{*}=\bm{a}^{t}\bm{b}=med$.
Hence, both the mediation effect defined by Equation \eqref{eq: nmediatineff}
and the indirect effect (i.e, $c'$) are invariant to the operator
of translation.

\subsubsection{Rotation}

Let $R$ be a $D$ by $D$ rotation matrix. In general, a rotation
matrix is an orthogonal matrix whose inverse and transpose matrices
are the same, and its determinant is 1. Let $R^{-1}$ and $R^{t}$
be the inverse and transpose matrices of $R$, respectively. Let ${\bf z}_{i}^{*}$
be the new position of actor $i$ after translation, i.e, ${\bf z}_{i}^{*}=R{\bf z}_{i}$
for actor $i$.

To study the association between the independent variables, new latent
positions, and the outcome variable, we fit the mediation model using
the new positions such that

\begin{equation}
\begin{cases}
R{\bf z}_{i} & =\bm{i}_{1}^{*}+\bm{a}^{*}X_{i}+\bm{\varepsilon}_{i1}^{*}\\
Y_{i} & =i_{2}^{*}+(\bm{b}^{*})^{t}R{\bf z}_{i}+c'^{*}X_{i}+\varepsilon_{i2}^{*}.
\end{cases}\label{eq:translation3}
\end{equation}
Thus 
\begin{equation}
\begin{cases}
{\bf z}_{i} & =R^{-1}\bm{i}_{1}^{*}+R^{-1}\bm{a}^{*}X_{i}+R^{-1}\bm{\varepsilon}_{i1}^{*}\\
Y_{i} & =i_{2}^{*}+(R^{t}\bm{b}^{*})^{t}{\bf z}_{i}+c'^{*}X_{i}+\varepsilon_{i2}^{*}.
\end{cases}\label{eq:translation4}
\end{equation}
By comparing the coefficients in Equation \eqref{eq:translation4}
with those in Equation \eqref{eq:network mediation 2}, we find, 
\begin{equation}
R^{-1}{\bm{a}}^{*}={\bm{a}}\text{ and }R^{t}{\bm{b}}^{*}={\bm{b}}\text{ and }c'^{*}=c
\end{equation}
or equivalently using the fact that $R^{-1}=R^{t}$, 
\begin{equation}
{\bm{a}}^{*}=R{\bm{a}}\text{ and }{\bm{b}}^{*}=R{\bm{b}}.
\end{equation}
The mediation effect of the new latent positions is 
\begin{equation}
med^{*}={{\bm{a}}^{*}}^{t}{\bm{b}}^{*}=(R{\bm{a}})^{t}(R{\bm{b}})={\bm{a}}^{t}R^{t}R{\bm{b}}={\bm{a}}^{t}{\bm{b}}=med
\end{equation}
using the fact that the rotation matrix $R$ is an orthogonal matrix
such that $R^{t}=R^{-1}$. Therefore, the network mediation effects
before and after the translation of the latent positions through rotation
are the same. The indirect effect also does not change.

\subsubsection{Reflection}

A reflection operator maps each position to a symmetry image about
some \emph{hyper-plane} in $\mathbb{R}^{D}$. The most imaginable
planes are those with one coordinate being 0. Without generality,
consider a plane formed by the first $D-1$ axes and all points on
this plane have $z_{D}=0$. For convenience, we name this hyper-plane
$P_{D}$. For a position ${\bf z}_{i}=(z_{i,1},z_{i,2},\cdots,z_{i,D})^{t}$,
its symmetric position about plane $P_{D}$ is ${\bf z}_{i}^{*}=(z_{i,1},z_{i,2},\cdots,z_{i,D-1},-z_{i,D})^{t}$.
Specifically, the first $D-1$ coordinates do not change, and the
last one reflects its sign.

When we fit the mediation model Equation \eqref{eq:network mediation 2}
to the new positions, the new coefficients ${\bm{a}}^{*}$, ${\bm{b}}^{*}$,
and $c'^{*}$ are the same as those obtained using the original latent
positions,

\begin{equation}
{\bm{a}}^{*}={\bm{a}}\text{ and }{\bm{b}}^{*}={\bm{b}}\text{ and }c'^{*}=c'.
\end{equation}
Thus, the mediation effect does not change before and after reflection.

For a general hyper-plane $P$, it can be transformed from $P_{D}$
through a series of translations and rotations, which do not influence
the the quantity ${\bm{a}}^{t}{\bm{b}}$ and $c'$. As a consequence,
the mediation effect defined as the inner product of coefficients
(Equation \ref{eq: nmediatineff}) and the direct effect $c'$ do
not change when the latent positions are reflected in general.

Based on the above articulation, both the mediation effect (Equation
\ref{eq: nmediatineff}) of a social network and the direct effect
are invariant to translation, rotation, and reflection of the latent
positions. Therefore, they are well-defined.

The total effect\footnote{The total effect is purely the sum of the indirect effect and direct
effect. It may not be the same as if regressing $Y$ on $X$ directly.
This is because the model complexity changes and the information used
in estimating the model is also different.} is defined as, 
\begin{equation}
c=c'+{\bm{a}}^{t}{\bm{b}}=c'+\sum_{d=1}^{D}a_{d}b_{d}.\label{nettotal}
\end{equation}
Throughout the rest of this article, the analysis will focus on the
mediation effect of a network, the indirect effect, and the total
effect.

\section{Model Estimation}

To estimate the network mediation effect, a Bayesian estimation method
is used, which is also used in the traditional mediation analysis
\citep{yuan2009bayesian,wang2011estimating,enders2013bayesian,wang2015moderated,miovcevic2018tutorial}.
The Bayesian inference on parameters $\bm{\theta}$ is based on the
posterior distribution given data $x$ and the priors distributions
of model parameters \citep{gelman2014bayesian,kruschke2014doing},
\begin{equation}
P(\bm{\theta}|x)\propto P(x|\bm{\theta})P(\bm{\theta}),
\end{equation}
where $P(\bm{\theta}|x)$ is the posterior distribution, $P(x|\bm{\theta})$
is the likelihood, and $P(\bm{\theta})$ is the joint prior distribution
of model parameters.

Assume the there are $N$ actors in the social network ${\bf M}$.
Given the Euclidean space $\mathbb{R}^{D}$, into which the actors
in a network mapped, the likelihood functions are listed below, 
\begin{align*}
{\bf z}_{i} & \sim\text{MVN}(\bm{i}_{1}+{\bm{a}}X_{i},\text{diag}(\sigma_{1,d}^{2},d=1,\cdots,D)),\\
\text{logit }(m_{ij}=1) & =\alpha-|{\bf z}_{i}-{\bf z}_{j}|,\\
Y_{i} & \sim\text{N}(i_{2}+c'X_{i}+{\bm{b}}^{t}{\bf z}_{i},\sigma_{2}^{2}).
\end{align*}

\textcolor{blue}{In the proposed mediation model, the unknowns include
model parameters $\bm{i}_{1},i_{2},\alpha,\bm{a},\bm{b},c',\sigma_{1,d}^{2}(d=1,2,\cdots,D),\sigma_{2}^{2}$.
In the currently analysis, we used the widely used priors for coefficients
and variance parameters. For the residual variance parameters of the
latent positions and the residual variance parameter of the outcome
variable $Y$, independent inverse Gamma (IG) priors are used. Specifically,
both $\sigma_{2}^{2}$ and $\sigma_{1,d}^{2}(d=1,2,\cdots,D$) follow
an inverse Gamma distribution with both the shape and position parameters
being set at the value 0.001. The regression coefficients $a_{d},b_{d}$
($d=1,\cdots,D$), the intercept parameter $i_{1,d}$ ($d=1,\cdots,D$),
$i_{2}$, the indirect effect parameter $c'$, and the slope parameter
$\alpha$ in the latent space model take independent normal priors
with the mean $0$ and the standard deviation $1000$. All those priors
are weakly informative. As in Bayesian structural equation modeling
\citep[SEM,][]{Depaoli2019,Lee2012,Muthen2012}, the prior
for the latent variable ${\bf z}_{i}$ is described by the model given
other parameter values. In the current study, the prior for ${\bf z}_{i}$
is $\text{MVN}(\bm{i}_{1}+{\bm{a}}X_{i},\text{diag}(\sigma_{1,d}^{2},d=1,\cdots,D))$
with given $\bm{i}_{1},$$\bm{a}$ and $(\sigma_{1,d}^{2},d=1,\cdots,D)$.
For each parameter, a chain of $20,000$ samples is drawn from its
conditional posterior distribution using the Gibbs sampler (Steps are provided in the Appendix). Samples for the network mediation effect and the total effect
parameters are computed using the samples of $a_{d},b_{d},c'(d=1,2,\cdots,D)$
through Equations \eqref{eq: nmediatineff} and \eqref{nettotal}.
For each chain, the burn-in period is 6,000 iterations and summary
statistics are computed based on the remaining part of the chain.
In both the simulation and empirical studies, the posterior means
computed using the posterior samples are the point estimates of model
parameters. The equal-tail $95\%$ credible intervals (CI) are reported
and they are used to evaluate whether the parameter estimates are
significant or not as done in the frequentist framework. }

\textcolor{blue}{The Bayesian estimation (BE) method accounts
for the potential dependence between the coefficient $\bm{a}$ and
$\bm{b}$. In classical mediation analysis, when structural equation
modeling is used to estimate and test the significance of mediation
analysis, the estimates of coefficient $a$ and $b$ are not independent
\citep{Kenny2018}. In the newly proposed network mediation analysis,
the dimensions of the social network are latent and the paths $\bm{a}$
and $\bm{b}$ could be correlated too. In the BE approach, we obtain
the posterior samples of $\bm{a}^{t}\bm{b}$ and the summary statistics based on those samples from the posterior distribution. Therefore,
the potential dependency between $\bm{a}$ and $\bm{b}$ are taken
into account in the posterior inference.}

\section{Illustration of Model Application: An Empirical Example}

\textcolor{blue}{The purpose of this section is to provide a step-by-step
illustration of how to use the newly proposed model to estimate the
mediation effect of social positions using social network data.} We
use the data collected by the Lab for Big Data Methodology at the
University of Notre Dame in 2017. The participants were 162 students
in a four-year college in China. There were 90 female and 72 male
students. Their average age was 21.64 years (SD=0.86). During the
data collection, each student indicated whether the other students
were his/her friends or not. We symmetrized the friendship network
using the strongest relation (i.e., if one of two students indicated
the other as a friend, we assumed they were mutual friends).

\begin{figure}
\begin{centering}
\includegraphics[scale=0.5]{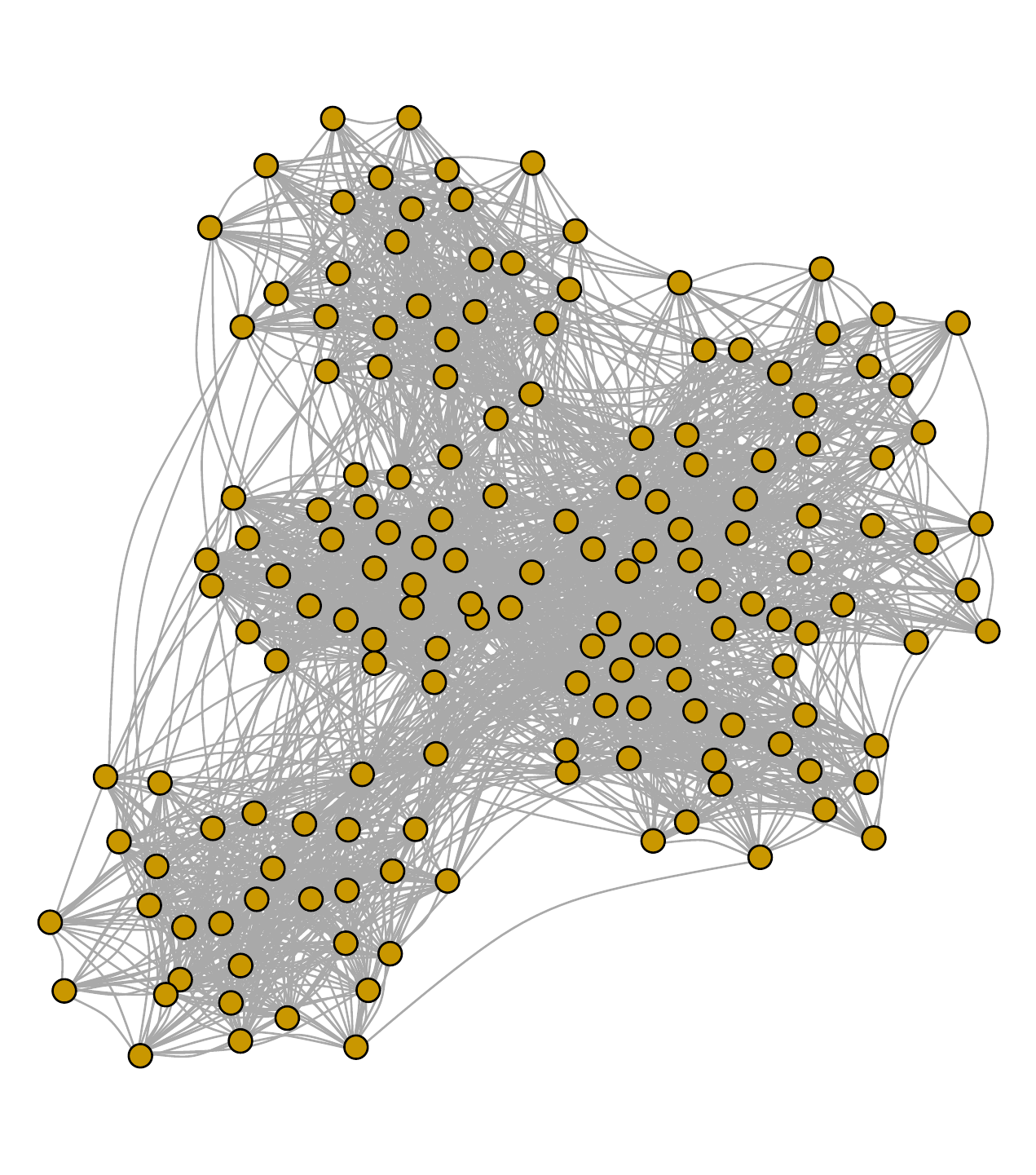} 
\par\end{centering}
\caption{The college friendship network with 162 students. Each dot represents
a student. A gray line between two dots indicates the existence of
friendship between two actors.\label{fig:friendship} }
\end{figure}

The data on the friendship were recorded in a 162 by $162$ adjacency
matrix ${\bf M}$, which was illustrated in Figure \ref{fig:friendship}.
Each dot represents a student and a gray line indicates the existence
of friendship between two students. The network density\footnote{The density of a social network is defined as the proportion of ties
over all possible ties.} is $16.2\%$. In addition to the information on the friendship, each
student also reported whether he/she smoked cigarettes or not retrospectively.
Among the 162 students, 43 students reported they had smoked cigarettes
during the past 30 days.

\textcolor{blue}{In the study, we are interested in how the gender
variable predicts students' social relations with others within a
friendship network and how such social relations relate to students'
smoking behaviors using social network data. Based on social theories,
gender affects patterns in social relations \citep{Fuhrer2002} and
social relations affect smoking \citep{Schaefer2013,Schane2009}.
Therefore, it is reasonable to study the potential mediation role
of a social network on the relationship between gender and smoking
behaviors of social actors. Although there are studies on the relationship
between gender and social relations and also studies on the impact
of social relations on smoking behavior, there is no study to investigate the  mediation role of a social network on the relationship between gender and smoking.}

\textcolor{blue}{To evaluate the indirect effect of
 ``gender'' on ``smoking behaviors'' through the friendship network, we fitted the newly proposed model for network
mediation analysis (Equations \ref{eq:network mediation 1} and \ref{eq:network mediation 2})
to the college friendship network data. }Because the outcome variable
$Y$ is binary (``0''=not smoking, ``1''=smoking), the latent
variable analysis was used assuming there was an underlying continuous
variable $Y^{*}$ such that the binary outcome variable is its dichotomy
with threshold 0,

\begin{equation}
\text{Mediation model: }\begin{cases}
{\bf z}_{i} & =\bm{i}_{1}+\bm{a}X_{i}+\bm{\varepsilon}_{i1}\\
Y_{i}^{*} & =i_{2}+\bm{b'}{\bf z}_{i}+c'X_{i}+\varepsilon_{i2}
\end{cases}\label{eq:network mediation 3}
\end{equation}
where $\varepsilon_{i2}$ follows the standard normal distribution.
The mediation effect is 
\[
med=\bm{a}^{t}\bm{b}=\sum_{d=1}^{D}a_{d}b_{d}
\]
with $D$ being the number of latent dimensions. The observed binary
variable $Y_{i}$ takes 1 if $Y^{*}\geq0$ and $Y_{i}=0$ if $Y^{*}<0$.
So the probability for $Y_{i}$ taking a value 1 conditional on its
social positions ${\bf z}_{i}$ and $X_{i}$ is $P(Y_{i}=1|{\bf z}_{i},X_{i})=\Phi(i_{2}+\bm{b}^{t}{\bf z}_{i}+c'X_{i})$. 

\begin{figure}[htbp]
\begin{centering}
\includegraphics[scale=0.5]{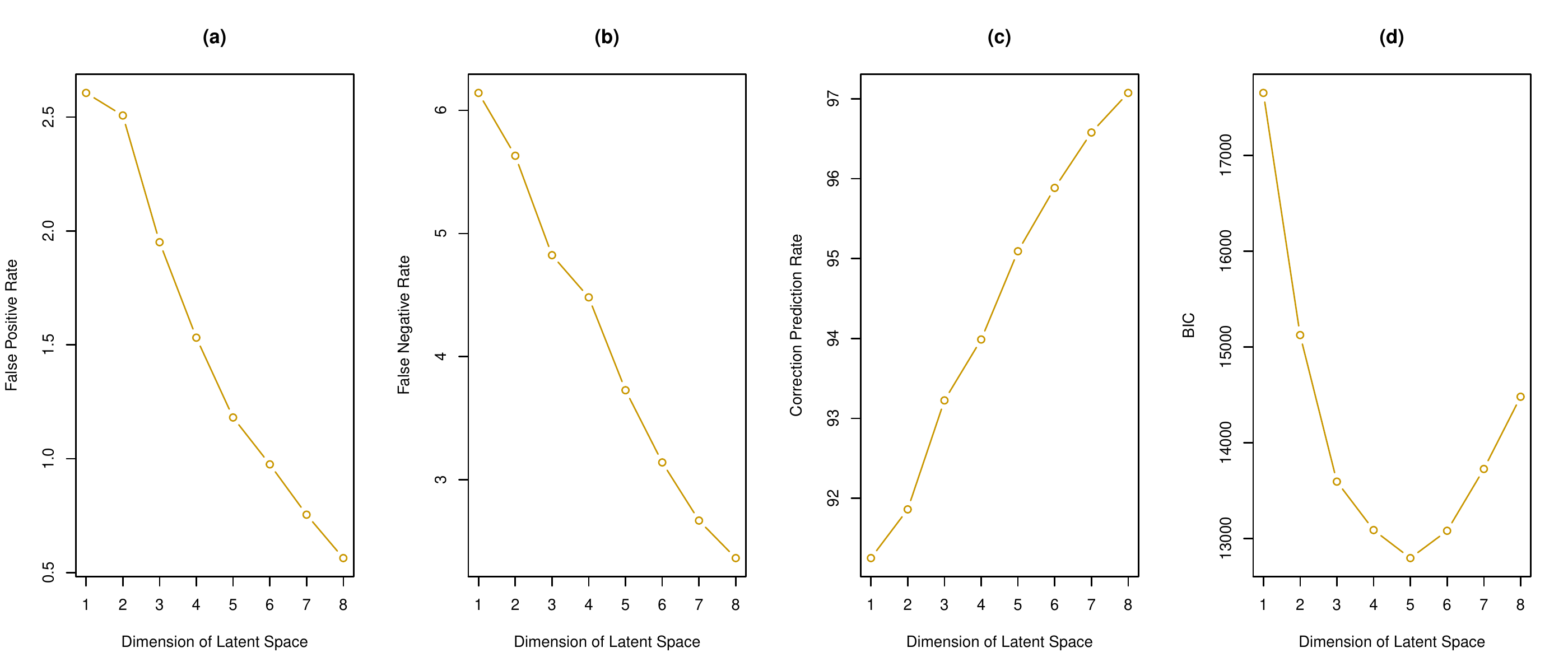} 
\par\end{centering}
\caption{(a) false positive rate, (b) false negative rate, (c) correct prediction
rate, (d) Bayesian Information Criterion (BIC) of latent space models
fit to the binary friendship network}
\label{fig:study3BIC} 
\end{figure}

In empirical studies, the first task is to determine the dimension
of the latent space before fitting the model to the data. According
to \citet{Hoff2002}, the axes of the latent space represent actors'
latent characteristics predicting the formation of manifest social
relations among actors. With more dimensions, more actor attributes
are used to explain the dependence of actors in a social network.
Since more information on actors is used to explain the dependence
among actors, a more accurate prediction of social relations would
be achieved. Meanwhile, the model complexity increases. As a result,
latent space models (Equation \ref{eq:network mediation 1}) were
fit to the binary friendship network in an exploratory manner by varying
the number of latent dimensions. The model was estimated using the
R package ``latentnet'' \citep{latentnet}. The Bayesian Information
Criterion \citep[BIC,][]{Schwarz1978} is used to help us to decide
the best number of latent dimension. The According to Figure \ref{fig:study3BIC}.
Based on plots (a), (b), and (c), the better prediction was achieved
with more latent dimensions. Meanwhile, the model parsimony suffers.
Considering also the model complexity, the optimal model was the one
with the smallest BIC. According to plot (d), the latent Euclidean
space was most plausible to have five dimensions. Thus the network
mediation model (Equations \ref{eq:network mediation 3}) with $D=5$
was fit to the friendship network and the model parameters estimates
are shown in Table \ref{tab:medparest}.

\textcolor{blue}{To understand the ``latent characteristics'' extracted
from the observed social network, we computed the Point-biserial correlation
coefficient (i.e., the the product-moment correlation coefficient
for a continuous variable and a dichotomous variable) of them with
``gender'' and ``smoking''. All the coefficients are provided
in Table \ref{tab:Point-biserial-correlation-of}. }

\textcolor{blue}{}
\begin{table}
\begin{centering}
\textcolor{blue}{}%
\begin{tabular}{ccccccc}
\hline 
 &  & \multicolumn{2}{c}{\textcolor{blue}{Gender}} &  & \multicolumn{2}{c}{\textcolor{blue}{Smoking}}\tabularnewline
\hline 
\textcolor{blue}{Latent Dimension} &  & \textcolor{blue}{cor} & \textcolor{blue}{p-value} &  & \textcolor{blue}{cor} & \textcolor{blue}{p-value}\tabularnewline
\hline 
\textcolor{blue}{${\bf z}_{1}$} &  & \textcolor{blue}{0.192} & \textcolor{blue}{0.015} &  & \textcolor{blue}{-0.114} & \textcolor{blue}{0.149}\tabularnewline
\textcolor{blue}{${\bf z}_{2}$} &  & \textcolor{blue}{-0.357} & \textcolor{blue}{$<0.001$} &  & \textcolor{blue}{0.215} & \textcolor{blue}{0.006}\tabularnewline
\textcolor{blue}{${\bf z}_{3}$} &  & \textcolor{blue}{-0.206} & \textcolor{blue}{0.009} &  & \textcolor{blue}{0.166} & \textcolor{blue}{0.034}\tabularnewline
\textcolor{blue}{${\bf z}_{4}$} &  & \textcolor{blue}{0.265} & \textcolor{blue}{0.001} &  & \textcolor{blue}{-0.363} & \textcolor{blue}{$<0.001$}\tabularnewline
\textcolor{blue}{${\bf z}_{5}$} &  & \textcolor{blue}{0.179} & \textcolor{blue}{0.023} &  & \textcolor{blue}{-0.075} & \textcolor{blue}{0.343}\tabularnewline
\hline 
\end{tabular}
\par\end{centering}
\textcolor{blue}{\caption{Point-biserial correlation of latent dimensions and gender and smoking\label{tab:Point-biserial-correlation-of}}
}
\end{table}
\textcolor{blue}{We found that the Point-biserial \cite[][]{Cheng2016} correlation of the
``latent dimensions'' with ``gender'' is
small to medium and they are all statistically significant. Some of
the latent dimensions are significantly linearly correlated with ``smoking''
but others are not. Therefore, a latent dimension is
a ``hybrid'' characteristic relevant to the friendship of a student
with others. All dimensions together describe the social positions
of actors in the friendship network.}

\textcolor{blue}{Considering the fact that the relationship between
gender and social positions are directional (i.e., gender influences
social positions) and the relationship between social relations and
smoking might be bi-directional, we thus fit a mediation model with
``gender'' as the independent variable and social network as the
mediator and ``smoking'' as the outcome variable for the purpose
of demonstration of how to fit the model to empirical data.} The estimated total effect (tot) was -2.17 with the equal-tail $95\%$
credible interval $[-2.982,-1.476]$ excluding 0. The estimated network
mediation effect (med) was -1.191 with the equal-tail $95\%$ credible
interval $[-2.254,-0.417]$. Similarly, the estimated direct effect
was -0.980 with the equal-tail $95\%$ credible interval $[-1.781,-0.065]$.
The result implies that female students smoked significantly less
than male students on average and gender influenced individuals' positions
in the friendship network and, in turn, the friendship network affected
individuals' smoking behaviors.

\begin{table}[!ht]
\centering{}\caption{Parameter estimates fitting the mediation model to the college friendship
network}
\label{tab:medparest} %
\begin{tabular}{ccccccc}
\hline 
D  & Par  &  & Est  &  & $2.5\%$  & $97.5\%$ \tabularnewline
\hline 
 & $c'$  &  & -0.980  &  & -1.781  & -0.065 \tabularnewline
5  & med  &  & -1.191  &  & -2.254  & -0.417 \tabularnewline
 & tot  &  & -2.17  &  & -2.982  & -1.476 \tabularnewline
\hline 
\end{tabular}
\end{table}
\textcolor{blue}{We would like to note that the focus of the empirical
study is not to build a new theory on causal relations between gender
and smoking, but to show practitioners how to implement the newly
proposed model for network mediation analysis. Although the $95\%$
credible interval of the mediation effect excludes 0, we still cannot
conclude decisively the mediation effect of a college friendship network on the
relationship between students' ``gender'' and ``smoking'' without further understanding the theoretical mechanism of the relationships among gender, friendship network, and smoking behaviors.}

\section{Simulation Study}

In this section, we will conduct a simulation study to investigate
how the Bayesian estimation method performs in estimating the mediation
effect of a social network. We are interested in how the performance
varies as the sample size and/or the model complexity (i.e., the number
of latent dimensions) change. In the following, we will first explain
the simulation design and the evaluation criteria and then present
the simulation results.

\subsection{Simulation design}

All the data sets are generated from the network mediation model in
Equations \eqref{eq:network mediation 1} and \eqref{eq:network mediation 2}.
In the simulation, the independent variable $X$ is generated from
the standard normal distribution, which is $X\sim\mbox{N}(0,1)$.
The dimensions of the latent space considered are $2$ and $3$.

To study the impact of the number of latent dimensions on the accuracy
of parameter estimates, we control the mediation effect to be the
same for $D=2$ and $3$. The coefficient $a_{d}=b_{d}$ along any
latent dimension $d$ takes values from $0$, $.14/\sqrt{D}$, $.39/\sqrt{D}$,
and $.59/\sqrt{D}$. This parameter specification guarantees that
the network mediation effect, i.e., $\sum_{d=1}^{D}a_{d}b_{d}$, is
the same even the number of latent dimensions varies. The corresponding
network mediation effects are thus $0$, $.0196$, $.1521$, and $.3481$
in the population model. For the direct effect parameter $c'$, two
values $c'=0.14$ and $c'=0,$ are considered, corresponding to the
partial and complete mediation, respectively.

All intercept parameters $(i_{1,1},\cdots,i_{1,D})^{t}$, $i_{2}$,
and $\alpha$ are set to be 0 in the data generating model. To make
both latent positions (${\bf z}=({z}_{1},\cdots,{z}_{D})^{t}$) and
the outcome variable ($Y$) to have unit variances, the residual variance
of latent factors (mediators) is computed by 
\begin{equation}
\sigma_{1,d}^{2}=var(\varepsilon_{i1,d})=1-a_{d}^{2}\hspace{1cm}\mbox{for }d=1,\cdots,D.
\end{equation}
And the residual variance of the outcome variable is calculated as
\begin{equation}
\sigma_{2}^{2}=var(\varepsilon_{2})=1-(\sum_{d=1}^{D}a_{d}b_{d}+c')^{2}-\sum_{d=1}^{D}b_{d}^{2}(1-a_{d}^{2}).
\end{equation}

It is important to note that both the latent position factors ${\bf z}_{d}(d=1,2,\cdots,D)$
and the dependent variable $Y$ have unit variances using the current
setup of variance parameters. Therefore, the network mediation effect
${\bm{a}}^{t}{\bm{b}}=\sum_{d=1}^{D}a_{d}b_{d}$ is the summation
of $D$ standardized indirect effects. Because the network mediation
effect is the summation of $D$ standardized indirect effects, it
is thus not a standardized effect. Because the sample size is closely
related to the performance of Bayesian estimation methods, we, therefore,
consider sample sizes 50, 100, 150, 200, 250, and 300 in the simulation
study.

It is worthy noting that with the above setup on the population parameters,
the mean of the squared distance (i.e., $d^{2}$) between two actors
is twice as large as the number of latent dimensions, i.e., $2D$.
Therefore, the distance of two actors in the latent space increases
statistically as $D$ increases. As such, the probability for two
actors to be connected decreases on average and the network density
thus decreases when the latent space has more dimensions in the population
model.

\begin{table}
\centering{}%
\begin{tabular}{lcr}
\hline 
Factors Considered  &  & Possible Values\tabularnewline
\hline 
Dimension of the latent space ($D$)  &  & 2, 3\tabularnewline
Sample size ($N$)  &  & 50, 100, 150, 200, 250, 300\tabularnewline
Network mediation effect (med) &  &  $0$, $.0196$, $.1521$, and $.3481$\tabularnewline
Indirect effect ($c'$)  &  & 0, .14\tabularnewline
\hline 
\end{tabular}\caption{Conditions manipulated in the simulation study\label{tab:Conditions-manipulated}}
\end{table}
The conditions considered in the simulation study are listed in Table
\ref{tab:Conditions-manipulated}. The factors we considered include
the number of dimensions in the latent space $D$, the sample size
$N$, the population network mediation effect $med$, and indirect
effect $c'$. Combining the levels of all the factors, there are $2\times6\times4\times2=96$
different conditions in total. For each condition, 500 data sets are
generated and model parameters for the network mediation model in
Equation \eqref{eq:network mediation 1} and \eqref{eq:network mediation 2}
are obtained using the Bayesian estimation method introduced in the
previous section.

\subsection{Evaluation criteria}

Bayesian parameter estimates are based on 14,000 Markov draws after
the burn-in phrase . The posterior mean based on the samples is computed
as 
\begin{equation}
\hat{\theta}=\frac{1}{14000}\sum_{i=6001}^{20000}\theta^{(i)}.
\end{equation}
Given a significance level $\alpha$, a posterior credible interval
of $r$th replication is defined as interval $[L_{r},R_{r}]$ such
that 
\begin{equation}
\frac{\#\{\theta^{(i)}:\theta^{(i)}<L_{r}\}}{14000}=\frac{\#\{\theta^{(i)}:\theta^{(i)}>R_{r}\}}{14000}=\alpha/2.
\end{equation}

Let $\theta$ be an arbitrary parameter in the model to be estimated
and also its population value. Let $\hat{\theta}_{r}$ and $[L_{r},R_{r}]$
be the posterior mean and $95\%$ credible interval from the $r$th
replication ($r=1,2,\cdots,500$). Let 
\begin{equation}
\bar{\theta}=\frac{1}{500}\sum_{r=1}^{500}\hat{\theta}_{r}.
\end{equation}
which is the average of parameter estimates across 500 replications.

The accuracy of parameter estimates is evaluated using ``relative
bias'', which is a ratio of bias (different between point estimate
and true value of a parameter) to the true value in percentage, 
\begin{eqnarray}
\text{relative bias}_{\theta} & =\begin{cases}
\frac{\bar{\theta}-\theta}{|\theta|}\times100\% & \text{ if }\theta\neq0\\
(\bar{\theta}-\theta)\times100\% & \text{ otherwise}.
\end{cases}
\end{eqnarray}

Moreover, the coverage probability of the Bayesian credible intervals
are also reported, which is the proportion of replications whose posterior
credible intervals cover the true parameter value $\theta$, 
\begin{equation}
\text{CR}_{\theta}=\frac{1}{500}\sum_{r=1}^{500}I(\theta\in[L_{r},R_{r}])\times100
\end{equation}
where $[L_{r},R_{r}]$ is the $95\%$ Bayesian credible interval.
The coverage rate is often used to assess the validity of the Bayesian
credible intervals. A coverage rate close to the nominal one (i.e.,
0.95) indicates the statistical inference based on the credible intervals
is trustworthy.

\subsection{Results}

The primary goal of a network mediation analysis is to assess the
indirect effect of a social network, the direct effect, and the total
effect of the independent variable(s). We thus record the estimates
of these three effects in each replication. The performance of the
Bayesian estimation method is examined from two aspects: (1) accuracy
of parameter estimates and(2) the coverage rates of Bayesian credible
intervals. The results of the simulation studies are summarized in
Figure \ref{fig:Bias123} and \ref{fig:CR123}. The plots are organized
into different panels based on the population network mediation effect,
the direct effect, and the number of dimensions of the latent space.

\subsubsection{Relative bias}

\begin{figure}
\begin{centering}
\includegraphics[scale=0.6]{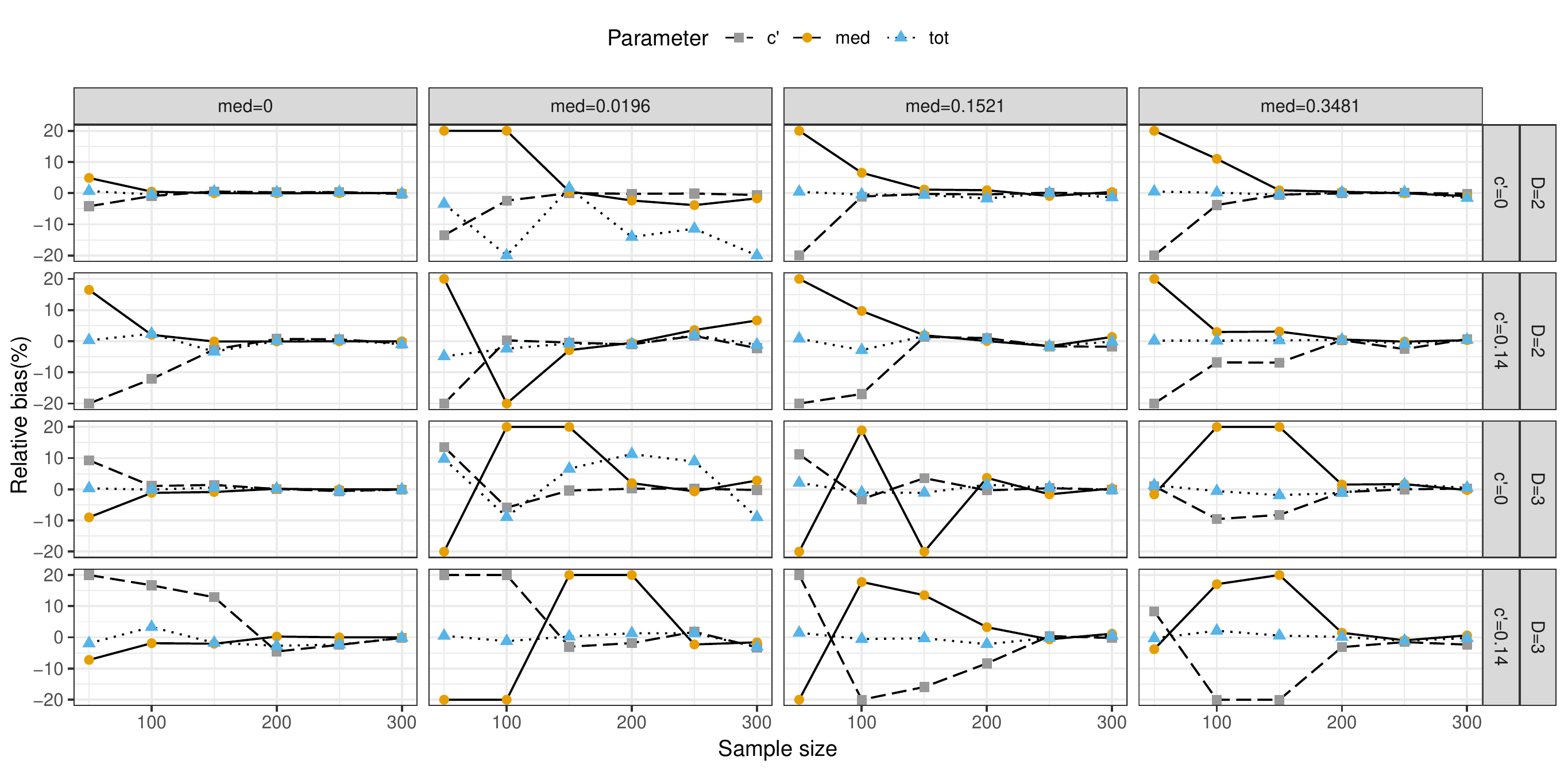} 
\par\end{centering}
\centering{}\caption{Relative bias of parameter estimates; ``med'' is the network mediation
effect; ``D'' is the number of dimensions of the latent social space;
`` $c'$'' is the direct effect\label{fig:Bias123}}
\end{figure}

The relative bias of the parameter estimates is shown in Figure \ref{fig:Bias123}.
For a case with the relative bias above $20\%$ or below $-20\%$,
we replace it with $20\%$ or $-20\%$ for better visualization. Therefore,
for a relative bias $20\%$ or $-20\%$ appearing in the plots in
Figure \ref{fig:Bias123}, its actual value should be more extreme
(larger than $20\%$ or less than $-20\%$). When the latent space
has two dimensions, the estimates of the mediation effect are biased
less than $5\%$ with a sample size 150 or larger. With a sample size
100, there are several cases with relative biases of the estimates
of the mediation effect larger than $10\%$. With a sample size 50,
all estimates of the mediation effect have relative biases larger
than $10\%$ except the condition with both the direct and mediation
effects being 0. Overall, a larger sample size leads to more accurate
parameter estimates. When the latent space has more dimensions, a
larger sample size is required to have comparable relative biases
in parameter estimates due to higher \textcolor{black}{model complexity.}

\subsubsection{Coverage rates}

\begin{figure}
\begin{centering}
\includegraphics[scale=0.6]{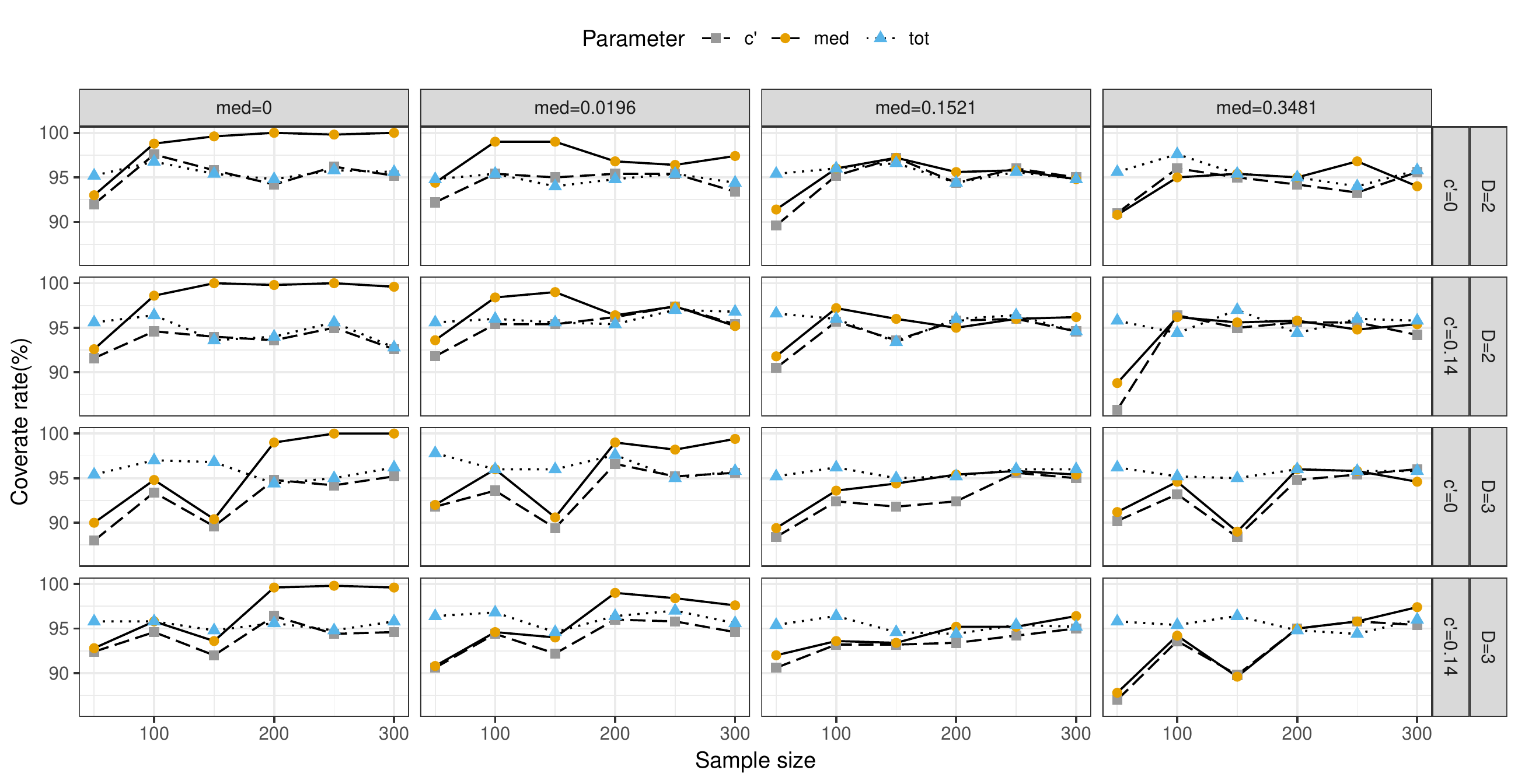} 
\par\end{centering}
\caption{Coverage rates of $95\%$ Bayesian credible intervals; ``med'' is
the network mediation effect; ``D'' is the number of dimensions
of the latent social space; `` $c'$'' is the direct effect.\label{fig:CR123}}
\end{figure}

\textcolor{black}{Coverage rates of the equal-tail $95\%$ Bayesian
credible intervals are provided in Figure \ref{fig:CR123}. The }reported
quantities are the coverage rates in percentage. In general, coverage
rates closer to $95\%$ are preferred. In the literature, a coverage
rate falling in the range $[92.5\%,97.5\%]$ is usually considered
to be acceptable. When the population mediation effect is 0, its coverage
rates are close to $100\%$ when the sample size is large compared
to the dimensions of the latent space, which is also found in traditional
mediation analysis \citep{yuan2009bayesian}. When the population
mediation effect is not zero, the coverage rates are acceptable when
the sample size is large enough for the given number of dimensions
in the latent space. For instance, when the latent space has two dimensions,
a sample size of at least 200 is required to have acceptable coverage
rates. When it has three dimensions, an even larger sample size is
required.

\textcolor{blue}{Overall, the coverage rates of the equal-tail $95\%$
CIs are consistent for conditions with a sample size above 200 for all parameters. However, when the sample size/network
size is small, the coverage rates for some parameters change as the
sample size varies. To further explore why the coverage rates vary, we looked into the condition with $med=.1521,c'=.14,D=2$.
From Figure \ref{fig:CR123}, we can notice that the coverage for
``med'' is lower at $N=50$ than $N=100$. It seems to contradict
the intuition that larger posterior variance with a smaller sample
size and thus a higher coverage rate of the posterior CIs. However,
we would like to show that it is plausible. For example, for the ``med''
parameter, the average width of the equal-tail $95\%$ CI is
2.252 and 0.572 with a sample size of $50$ and $100$ respectively.
The CI is indeed wider with a small $N$. Posterior mean is used as the Bayesian estimator in the current study (both empirical study and simulation study) and
estimates based on samples after the burn-in period from the posterior
distribution is used as the parameter estimates for each replication.
However, the relative bias for ``med'' is much larger with $N=50$
than $100$ as shown in Figure \ref{fig:Bias123}. It indicates that
the ``center'' (not exactly the middle of CI) of CIs deviates further
away from the true parameter value. Although the CI is wider, the
entire CI is further away from the true value, which leads to low
coverage rates.} \textcolor{blue}{This is also confirmed by the plots of CIs with
$N=50$ and $100$, for which we randomly sample 100 from the total
500 replications without replacement for the ease of plots. On the
left panel, there are 100 CIs with $N=50$ and on the right panel,
there are also 100 CIs with $N=100$. CIs in the left plots are wider
than those in the right plot, which is what we have expected. However,
more cases are failing to cover the true value (i.e., blue line) in
the left plot.}

\begin{figure}
\begin{centering}
\includegraphics[scale=0.45]{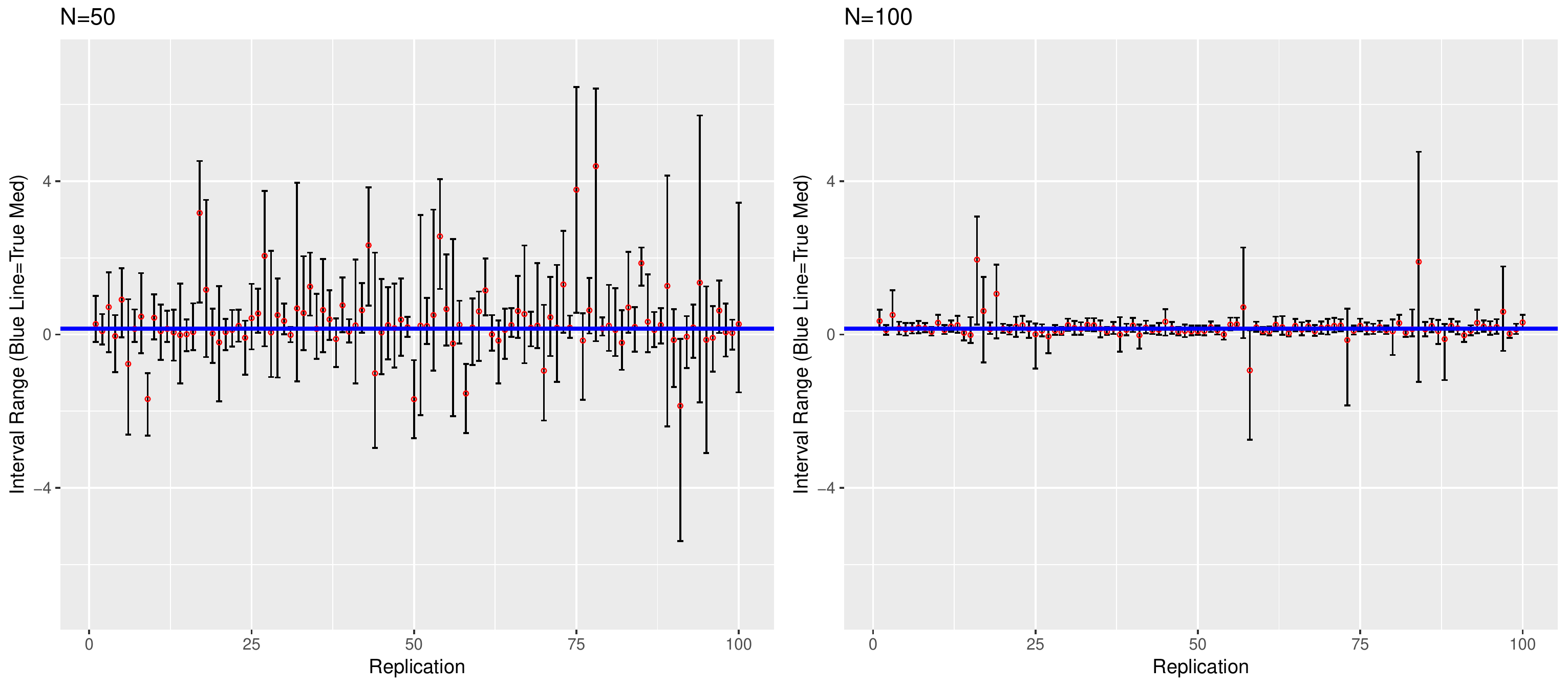}
\par\end{centering}
\caption{\textcolor{blue}{Equal-tail $95\%$ credible intervals for the parameter
``med'', the blue line is the mediation effect in the population
model and the red dots are the posterior means for each replication}}
\end{figure}

\section{Discussion}

The purpose of this study was to develop a model and a Bayesian estimation
method to estimate the mediation effect of a social network on the
relationship between an independent and a dependent variable. Social
network data are, however, high-dimensional with $O(N^{2})$ observations
on the dyadic variable and lack dependence (e.g., dyads sharing an
actor depend on each other), which violates the assumptions of many
statistical modeling techniques. 

To deal with the challenges posed by the uncommon features of social
network data, we used the latent space modeling to find a lower-dimension
representation of social networks. Latent space modeling maps actors
to positions in a latent space formed by actor's latent characteristics.
The distance of two actors in that latent space quantifies their (dis)similarity
in those latent characteristics and predicts the propensity for them
to be connected in the manifest social network. The position of an
actor represents his/her latent characteristics. Such characteristics
may be predicted by an independent variable (e.g., gender) and predict
the dependent variable (e.g., smoking behavior). Therefore, they act
as the actual mediators in analysis. Because the latent space is invariant
to the operator of rotation, translation, and projection, actors'
latent positions are not uniquely determined with constrains on latent
positions. However, we showed that the proposed network mediation
effect is still well-defined regardless of the indeterminacy of the
latent positions.

\textcolor{blue}{The proposed model could be used to evaluate and
test the hypothesized mediation effect when one already knows there is a causal path from $X$ to the social network to $Y$ based
on social and behavioral theories \citep{Sweet2019,mackinnon2012introduction}.
To use the model, there should be no unmeasured confounding of the
relationship between $X$, the social network, and $Y$. To make the
defined quantity $\bm{a}^{t}\bm{b}$ to represent the indirect effect,
the relationship between variables should be linear and there should
also be no directional path across latent dimensions. When a social
network is a mediator based on theory, our newly proposed model and
estimation method can evaluate and test the mediation
effect. }

To estimate the proposed model, we adopted a Bayesian estimation method.
\textcolor{blue}{Posterior inference on the indirect effect $\bm{a}^{t}\bm{b}$
is obtained based on the samples from its posterior distribution and
it, therefore, accounts for the potential dependence of the coefficient
$\bm{a}$ and $\bm{b}$}. A simulation study was conducted to evaluate
the performance of the Bayesian method in estimating the network mediation
effect. According to the results of the simulation study, the Bayesian
estimation method could provide accurate parameter estimates. With
more dimensions in the latent space, more instances with biases larger
$10\%$ occurred for a given small sample size such as 50 or 100.
When the sample size is above 100, the parameter estimates were generally
accurate. The coverage rates of the equal-tail $95\%$ credible intervals
are mostly in the acceptable range $[92.5\%,97.5\%]$ with a few exceptions
with the true mediation effect being 0, which was also observed in
the simple mediation analysis as discussed in \citet{yuan2009bayesian}.

To illustrate the application of the proposed network mediation model,
we analyzed a college friendship network. We found that female students
smoked cigarettes significantly less than male students on average.
A part of the gender difference in smoking behaviors was explained
by the friendship network. Hence, the students' gender influenced
their friendship with others within a network. and in turn, the friendship
network affected the students' smoking behaviors. Furthermore, the
students' gender directly influenced their smoking behaviors.

In the empirical data analysis, we fit a model in which the friendship
network predicts the students' smoking behaviors. However, the relationship
between the friendship network and the smoking behaviors might be
bi-directional. Smoking behaviors may also influence the students'
social relations with others. To address the potential bi-directional
relations, we will extend our model to longitudinal network mediation
analysis in the future. Although the proposed network mediation model
was presented based on binary networks, our modeling framework can
be easily extended to valued networks with ordinal social relations.
To do that, we only need to change the link function in the latent
space modeling. It can also be modified to model directed networks
(i.e., the relations are asymmetry), for which we need to use a properly
defined ``distance'' as \citet{Hoff2002}.

\textcolor{blue}{This study is not without limitations. In the current
study, we proposed to evaluate the indirect effect of the entire social
network. We lack the information to interpret the indirect effect
along a single latent dimension. This is because we cannot name the
axes of the latent space with only network data. To better understand the latent
dimensions, we need additional information for each latent dimension.
In the future, we may include indicators for each latent dimension
as being done in exploratory factor analysis \citep{Cattell1952}. }

\textcolor{blue}{As discussed by \citet{Shalizi2013}, the generalization
of statistical inference from a sub-network to the larger network
is only valid for very special model specifications. A necessary condition
for the generalizability of network inference is that a network must
be $\emph{projective}$ \citep{Shalizi2013}, which means that the
sufficient statistics of it have independent and separate increments
when the network has more actors. Most popular specifications for
social networks including the latent space models and stochastic graph
models cannot be projective \citep{Shalizi2013}. That means that
the out-of-sample prediction is not plausible in social network analysis
in general and the proposed mediation analysis is not an exception.}

\textcolor{blue}{The effect of a latent space on the probability of
having edges between nodes is heavily influenced by the definition
of the latent space. In our future research, we plan to develop other
network mediation analysis frameworks by adopting other specifications
in latent effects - for example, latent spaces based on principles
of eigenanalysis \citep{Hoff2008}. There are also many other ways
to describe the social position of an actor in a social network. For
instance, we can use the centrality measures, e.g., the number of
friends an actor has, as the actual mediator. In the current model,
the variables are assumed to follow the linear relationship. The modeling
framework can be extended to nonlinear relations. And it can also
be extended to the longitudinal frameworks \citep{Jose2016,Roth2012}
and causal inferences \citep{VanderWeele2015}. }

\bibliographystyle{rss}
\bibliography{LSM}
\newpage

\subsection{Gibbs sampler}

Because the posterior distribution has no closed form, we thus use
the Markov Chain Monte Carlo method to draw samples of parameters
from their posterior distributions. The steps of Gibbs sampler is
provided below. Given desired length $T$ and initials ($\bm{i}_{1}^{0},i_{2}^{0},\bm{a}^{0},\bm{b}^{0}$,$\alpha^{0},c'^{0},$$(\sigma_{1,d}^{2},d=1,2,\cdots,D)^{0}$,$(\sigma_{2}^{2})^{0}$),
\begin{enumerate}
\item In the $k'$th iteration, draw ${\bf z}_{i}^{k}$ from its conditional
posterior distribution $P({\bf z}_{i}|\bm{i}_{1}^{k-1},i_{2}^{k-1},\bm{a}^{k-1},\bm{b}^{k-1},\alpha^{k-1},(c'){}^{k-1},(\sigma_{1,d}^{2},d=1,2,\cdots,D)^{k-1},(\sigma_{2}^{2})^{k-1},X, Y,{\bf M})$,
for all actors $i$ in the network;
\item Draw $\bm{i}_{1}^{k}$ from its conditional posterior distribution
$P(\bm{i}_{1}|i_{2}^{k-1},\bm{a}^{k-1},\bm{b}^{k-1},\alpha^{k-1},(c')^{k-1},(\sigma_{1,d}^{2},d=1,2,\cdots,D)^{k-1},(\sigma_{2}^{2})^{k-1},{\bf z}_{i}^{k},X,Y,{\bf M})$
with updated ${\bf z}_{i}^{k};$
\item Draw $i_{2}^{k}$ from its conditional posterior distribution $P(i_{2}|\bm{i}_{1}^{k},\bm{a}^{k-1},\bm{b}^{k-1},\alpha^{k-1},c'^{k-1},(\sigma_{1,d}^{2},d=1,2,\cdots,D)^{k-1},(\sigma_{2}^{2}){}^{k-1},{\bf z}_{i}^{k},X, Y,{\bf M})$
\item Draw $\alpha^{k}$ from its conditional posterior distribution $P(\alpha|\bm{i}_{1}^{k},i_{2}^{k},\bm{a}^{k-1},\bm{b}^{k-1},(c'){}^{k-1},(\sigma_{1,d}^{2},d=1,2,\cdots,D)^{k-1},(\sigma_{2}^{2})^{k-1},\text{{\bf z}}_{i}^{k},X, Y,{\bf M}$);
\item Draw $\bm{a}^{k}$from its conditional posterior distribution $P$($\bm{a}|\bm{i}_{1}^{k},i_{2}^{k},\bm{b}^{k-1},\alpha^{k},(c'){}^{k-1},(\sigma_{1,d}^{2},d=1,2,\cdots,D)^{k-1},(\sigma_{2}^{2})^{k-1}$,${\bf z}_{i}^{k},X, Y$, ${\bf M}$);
\item Draw $\bm{b}^{k}$ from its conditional posterior distribution $P$($\bm{b}|\bm{i}_{1}^{k},i_{2}^{k},\bm{a}^{k},\alpha^{k},(c')^{k-1},(\sigma_{1,d}^{2},d=1,2,\cdots,D)^{k-1},(\sigma_{2}^{2})^{k-1}$,${\bf z}_{i}^{k},X,Y$, ${\bf M}$);;
\item Draw $c'^{k}$ from its conditional posterior distribution $P$($c'|\bm{i}_{1}^{k},i_{2}^{k},\bm{a}^{k},\text{\ensuremath{\bm{b}^{k}}},\alpha^{k},(\sigma_{1,d}^{2},d=1,2,\cdots,D)^{k-1},(\sigma_{2}^{2})^{k-1}$,${\bf z}_{i}^{k},X,Y$,${\bf M}$);;
\item Draw $(\sigma_{1,d}^{2},d=1,2,\cdots,D)^{k}$ from its conditional
posterior distribution $P$($\sigma_{1,d}^{2},d=1,2,\cdots,D|\bm{i}_{1}^{k},i_{2}^{k},\bm{a}^{k},\text{\ensuremath{\bm{b}^{k}}},\alpha^{k},c'^{k},(\sigma_{2}^{2})^{k-1}$,${\bf z}_{i}^{k},X,$${\bf M}$);
\item Draw $(\sigma_{2}^{2})^{k}$ from its conditional posterior distribution
$P$($\sigma_{2,}^{2}|\bm{i}_{1}^{k},i_{2}^{k},\bm{a}^{k},\text{\ensuremath{\bm{b}^{k}}},\alpha^{k}$,$(\sigma_{1,d}^{2},d=1,2,\cdots,D)^{k},X,Y$,${\bf M}$).
\end{enumerate}
Repeat steps 1-9 until the chain reach convergence and has sufficient
posterior samples.

\subsection{OpenBUG Code}

The following is the OpenBUG code used for empirical data analysis with three latent dimensions. 
\begin{verbatim}
model<-function(){      
    for( i in 1: N){       
        for(k in 1:3){        
       z[i, k]~dnorm(mu1[i, k], pre[k])        
       mu1[i, k]<-(i1[k]+a[k]*x[i])}
       y[i]~dcat(pi[i, 1:2])      
       pi[i,1] <-phi(-i2-bc[i])    
       bc[i]<-b[1]*z[i,1]+b[2]*z[i,2]+b[3]*z[i,3]+c*x[i]    
      pi[i,2] <-1-phi(-i2-bc[i])      }     
#latent space model" upper triangle    
   for (i in 1:(N-1)){          
       for (j in (i+1):N){           
          m[i,j] ~ dbern(p[i,j])              
         for(k in 1:3){                 
            diff[i,j,k] <- (z[i, k] - z[j, k]) }        
            d[i,j] <- inprod(diff[i,j,1:3], diff[i,j,1:3])      
           logit(p[i,j]) <- alpha -sqrt(d[i,j]) }        }  
        #prior      
        alpha~dnorm(0, 0.001)      
        i2~dnorm(0, 0.001)      
        c~dnorm(0, 0.001)     
      for(k in 1:3){        
        a[k]~dnorm(0, 0.001)       
        b[k]~dnorm(0, 0.001)     
        i1[k]~dnorm(0, 0.001)            
        pre[k]~dgamma(0.001,0.001)                 }          
        sig11<-1/pre[1]       
        sig12<-1/pre[2]        
        sig13<-1/pre[3] 
          
       med<-inprod(a[1:3], b[1:3])      
      tot<-med+c   } 
#initial
list(i1=c(0,0,0),i2=0,a=c(0,0,0),b=c(0,0,0),c=0,pre=c(1,1,1))
#data 
list(N, X, Y, M)
\end{verbatim}

\end{document}